\documentclass[aps,graphicx,prb,twocolumn,groupedaddress]{revtex4}
\usepackage[dvips]{graphics}
\usepackage[dvips]{graphicx}
\usepackage[]{amsmath}
\usepackage[]{amssymb}
\usepackage[]{amsfonts}

\newcommand{\ket}[1]{\left|#1\right>}
\newcommand{\bra}[1]{\left<#1\right|}

\begin{document}
\title{Ultrashort Lifetime Expansion for Indirect Resonant Inelastic X-ray Scattering}
\author{Luuk J. P. Ament$^1$, Filomena Forte$^{1,2}$ and Jeroen van den Brink$^{1,3}$}
\affiliation{$^1$ Institute-Lorentz for Theoretical Physics,  Universiteit  Leiden,\\
P.O. Box 9506, 2300 RA Leiden,The Netherlands\\
$^2$ Dipartimento di Fisica "E. R. Caianiello", Universit{\`a} di Salerno, I-84081 Baronissi, Salerno, Italy and Laboratorio Regionale SuperMat, INFM-CNR, Baronissi (SA), Italy\\
$^3$ Institute for Molecules and Materials, Radboud Universiteit Nijmegen,\\ 
P.O. Box 9010, 6500 GL Nijmegen, The Netherlands}
\date{\today}

\begin{abstract}
In indirect resonant inelastic X-ray scattering (RIXS) an intermediate state is created with a core-hole that has an ultrashort lifetime. The core-hole potential therefore acts as a femtosecond pulse on the valence electrons. We show that this fact can be exploited to integrate out the intermediate states from the expressions for the scattering cross section. By this we obtain an effective scattering cross section that only contains the initial and final scattering states. We derive in detail the effective cross section which turns out to be a resonant scattering factor times a linear combination of the charge response function $S({\bf q},\omega)$ and the dynamic longitudinal spin density correlation function. This result is asymptotically exact for both strong and weak local core-hole potentials and ultrashort lifetimes. The resonant scattering pre-factor is shown to be weakly temperature dependent. We also derive a sum-rule for the total scattering intensity and generalize the results to multi-band systems. One of the remarkable outcomes is that one can change the relative charge and spin contribution to the inelastic spectral weight by varying the incident photon energy.
\end{abstract}

\maketitle
\narrowtext

\section{Introduction}
Resonant Inelastic X-ray Scattering (RIXS) is a technique that matures rapidly due to the recent increase in brilliance of the new generation synchrotron X-ray sources, where high flux photon beams with energies that are tunable to resonant edges are now becoming widely available~\cite{Kotani01}. The probability for X-rays to be scattered from a solid state system can be enhanced by orders of magnitude when the energy of the incoming photons is in the vicinity of an electronic eigenmode --a resonant edge-- of the system. RIXS experiments are performed on  e.g. the K-edges of transition metal ions, where the frequency of the X-rays is tuned to match the energy of an atomic $1s$-$4p$  transition, which is around 5-10 keV~\cite{Hasan00,Kim02,Hill98,Isaacs96,Kao96,Inami03,Abbamonte99,Doering04,Suga05,Nomura05,Wakimoto05,Collart06,Seo06}. At this resonant energy a $1s$ electron from the inner atomic core is excited into an empty $4p$ state, see Fig. 1. 

It is a well-known fact that the $1s$ core-hole that is created has an ultrashort lifetime, of the order of femtoseconds. The reason is that the core-hole has a very high energy and is prone to decay via all sorts of radiative and non-radiative processes, severely cutting down the efficiency of RIXS. In theoretical treatments of RIXS this life-time effect is normally introduced as a core-hole broadening and disregarded from that point on. 

In a previous study~\cite{Brink03}, however, we have shown that from the theory perspective there is a great advantage of the very short lifetime of the core-hole. The ultrashort lifetime implies that for the other electrons in the system --particularly for the slow ones that are close to the Fermi-energy-- the core-hole potential is almost an instantaneous delta-function in time. This allows for a systematic expansion of the scattering cross section in terms of the lifetime, for which we present a detailed derivation and various generalizations in this paper. We shall see that the most important consequence of the ultrashort core-hole lifetime is that for indirect RIXS the effective scattering cross section is proportional to the charge structurefactor $S({\bf q},\omega)$ and the longitudinal spin structure factor that is associated with it.

The indirect RIXS process is shown schematically in Fig. 1. In transition metal systems the  photo-electron is promoted from a $1s$ core-orbital to empty $4p$ states that are far (10-20 eV) above the Fermi-level. So the X-rays do not cause direct transitions of the $1s$ electron into the lowest $3d$-like  conduction bands of the system. Still RIXS is sensitive to excitations of electrons near the Fermi-level. The Coulomb potential of $1s$ core-hole causes e.g. very low energy electron-hole excitations in the valence/conduction band: the core-hole potential is screened by the valence electrons.  When the excited $4p$-electron recombines with the $1s$ core-hole and the outgoing photon is emitted, the system can therefore be left behind in an excited final state.  Experimentally the momentum ${\bf q}$ and energy $\omega$ of the elementary excitation is determined  from the difference in energy and momentum between incoming and outgoing photons. Since the excitations are caused by the core-hole, we refer to this scattering mechanism as {\it indirect} resonant inelastic X-ray scattering (RIXS).

At present energy resolutions of about 100 meV can be reached. In the near feature it seems experimentally feasible for RIXS to become sensitive to the low energy excitations of the solid, where excitation energies are of the order of room temperature.  Recently it has been shown that also magnetic excitations, magnons, can be measured in RIXS~\cite{Hill_tbp,Brink05}. Other interesting low-lying electronic excitations that potentially can be probes by RIXS are, for example, collective features such as plasmons, orbitons, excitons, but also single-particle-like continua related to the band structure. RIXS provides a new tool to study these elementary excitations.

For the interpretation of spectroscopic data, it is very important to express the scattering cross section for a technique in terms of physical correlation functions. In this paper, we derive in detail the dynamical correlation function that is measured in indirect resonant inelastic X-ray scattering. For local core-hole potentials and ultrashort lifetimes, the dynamical correlation function turns out to be a linear combination of the charge density and longitudinal spin density response function. We show that for a single band system the actual linear combination that is measured depends on the energy of the incoming photons and we determine the precise energy dependence of its coefficients. A sum-rule is derived and we generalize these results to the case of finite temperature and for multiband systems. 

\begin{figure}
\includegraphics[width=0.75\columnwidth]{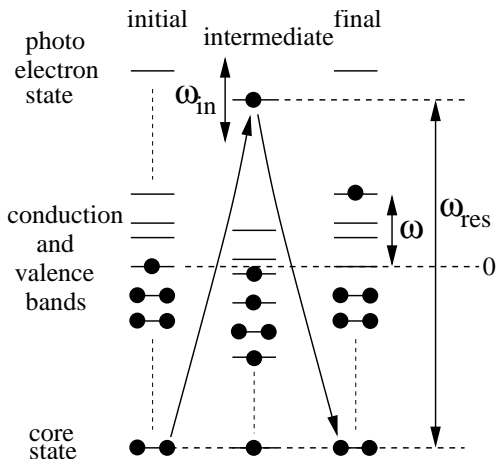}
\caption{Schematic representation of the indirect resonant inelastic X-ray scattering (RIXS) process.}
\label{fig:scheme}
\end{figure}

\section{Series expansion of the scattering cross section}
The Kramers-Heisenberg formula~\cite{Kramers25,Platzman69,Klein83,Blume85} for the resonant X-ray scattering cross section at finite temperature is
\begin{equation}
\left. \frac{ {\rm d}^2 \sigma} { {\rm d} \Omega {\rm d} \omega}
\right|_{res} \propto \left\langle\sum_{\textsc{f}}  |
A_{\textsc{fi}} |^2 \ \delta (\omega-\omega_{\textsc{fi}})
\right\rangle_{\rm T}, \label{eq:Kramers}
\end{equation}
where $\textsc{f}$ and $\textsc{i}$ denote the final and initial
state of the system, respectively. The sum is over all final
states and the brackets denote the statistical average over
initial states $\textsc{i}$ for a temperature $T$.
The momentum and energy of the incoming/outgoing photons is ${\bf
q}_{{\rm in}/{\rm out}}$ and $\omega^0_{{\rm in}/{\rm out}}$ and
the loss energy $\omega= \omega^0_{\rm out}-\omega^0_{\rm in}$ is
equal to the energy difference between the final and initial state
$\omega_{\textsc{fi}}=E_\textsc{f}-E_\textsc{i}$.  In the
following we will take the groundstate energy of our system as
reference energy: $E_{\rm gs}\equiv 0$. The scattering amplitude
$A_{\textsc{fi}}$ is given by
\begin{eqnarray}
A_{\textsc{fi}} = \omega_{\rm res} \sum_{ n} \frac{\langle
\textsc{f} | \hat{O} | n \rangle \langle n | \hat{O} | \textsc{i}
\rangle } {\omega_{\rm in}-E_n -i \Gamma}, \label{eq:amplitude}
\end{eqnarray}
where $\omega_{\rm res}$ the resonant energy, $n$ denotes the intermediate states and
$\hat{O}$ the (dimensionless) dipole operator that describes the excitation from initial
to intermediate state and the de-excitation from intermediate to final state.
The energy of the incoming X-rays with respect to the resonant energy is $\omega_{\rm in}$ (this energy can thus
either be negative or positive: $\omega_{\rm in}= \omega^0_{\rm in}-\omega_{\rm res}$) and $E_n$ is the
energy of intermediate state $|n\rangle$ with respect to the resonance energy.

In the intermediate state a core-hole and a photo-excited electron are present.
When we take the Coulomb interaction between the intermediate state core-hole and the valence band electrons
into account, we obtain a finite inelastic scattering amplitude. In that case there is a non-zero probability that
an electron-hole excitation is present in the final state, see Fig.1.


The intermediate state, however, is not a steady state. The highly energetic $1s$ core-hole quickly decays e.g.
via Auger processes and the core-hole life-time is very short. The Heisenberg time-energy
uncertainty relationship then implies that the core-hole energy has an appreciable uncertainty. This
uncertainty appears in the formalism above as the core-hole energy broadening $\Gamma$ which is
proportional to the inverse core-hole life-time, which is of the order of electron volts as the lifetime is ultrashort, of the order of femtoseconds.
%
%
Note that the life-time broadening only appears in the intermediate states and not in the final or initial
states as these both have very long life times. This implies that the core-hole broadening does not present an
intrinsic limit to the experimental resolution of RIXS : the loss energy $\omega$ is completely determined by kinematics.

When the incoming energy of the X-rays is equal to a resonant
energy of the system $\omega_{in} - E_n =0$ and we see from
Eqs.~(\ref{eq:Kramers},\ref{eq:amplitude}) that the resonant
enhancement of the X-ray scattering cross section is $(\omega_{\rm
res}/\Gamma)^2$, which is $\sim 10^6$ for a transition metal
K-edge~\cite{Blume85}.

In a resonant scattering process, the measured system  is generally strongly perturbed. Formally this is clear from the Kramers-Heisenberg formula~(\ref{eq:Kramers}), in which  both the energy and the wavefunction of the intermediate state --where a potentially strongly perturbing core-hole is present-- appear. This is in contrast with canonical optical/electron energy loss experiments, where the probing photon/electron presents a weak perturbation to the system that is to be measured.

To calculate RIXS amplitudes, one possibility is to numerically evaluate the Kramers-Heisenberg  expression. To do so, all initial, intermediate and final state energies and wavefunctions need to be known exactly, so that in practice a direct evaluation is only possible for systems that, for example, consist  of a small cluster of atoms~\cite{Tohyama02}.
In this paper, however, we show that under the appropriate conditions we can integrate out the intermediate states from the Kramers-Heisenberg expression. After doing so, we can directly relate RIXS amplitudes to linear charge and spin response functions of the unperturbed system. For non-resonant scattering, one is familiar with the situation that the scattering intensity is proportional to a linear response function, but for a resonant scattering experiment this is a quite unexpected result.

Let us proceed by formally expanding the scattering amplitude in a power series
\begin{eqnarray}
A_{\textsc{fi}} = \frac{ \omega_{\rm res} } {\omega_{\rm in} - i
\Gamma} \sum_{l=0}^{\infty} M_l, \label{eq:A_fi}
\end{eqnarray}
where we introduced the matrix elements
\begin{eqnarray}
M_l= \sum_{ n} \left( \frac{E_n}{\omega_{\rm in}-i\Gamma}
\right)^l \langle \textsc{f} | \hat{O} | n \rangle \langle n |
\hat{O} | \textsc{i} \rangle. \label{eq:M_l}
\end{eqnarray}
The formal radius of convergence of this power series is given by $E^2_n/(\omega_{\rm in}^2+\Gamma^2)$, so that
the series is obviously convergent when the incoming X-ray energy is e.g far enough below the resonance,
i.e. when $|\omega_{\rm in}| \gg 0 $. But also at resonance, when $\omega_{\rm in} = 0 $ the series is convergent
for intermediate energies that are smaller than the core-hole broadening $\Gamma$. Thus this expansion is controlled for ultrashort core-hole lifetimes, which implies that $\Gamma$ is large. In the following we will be performing  re-summations of this series.

We denote the denominator of the expansion parameter $\omega_{\rm in}-i\Gamma$ by the complex number
$\Delta$, so that
\begin{eqnarray}
M_l &=& \frac{1}{\Delta^l}
\sum_{ n} \langle \textsc{f} | \hat{O}| n \rangle (E_n)^l \langle n | \hat{O} | \textsc{i} \rangle \nonumber \\
    &=& \frac{1}{\Delta^l} \langle \textsc{f} | \hat{O} (H_{\rm int})^l \hat{O} | \textsc{i} \rangle,
\label{eq:Mterms}
\end{eqnarray}
where $H_{\rm int}$ is the Hamiltonian in the intermediate state. We thus obtain the following series expansion for the resonant cross section
\begin{eqnarray}
\left. \frac{ {\rm d}^2 \sigma} { {\rm d} \Omega {\rm d} \omega}
\right|_{res}  \propto \left\langle \sum_{\textsc{f}}  \left|
\frac{\omega_{\rm res}}{\Delta} \sum_{l=0}^{\infty} M_l \right|^2
\delta (\omega-\omega_{\textsc{fi}})\right\rangle_{\rm T}.
\label{eq:exp_cross}
\end{eqnarray}

\section{Indirect RIXS for spinless fermions: T=0}
We will first calculate the resonant X-ray cross section at zero temperature in
the case where the valence and conduction electrons are
effectively described by a single band of spinless fermions: spin, and
orbital degrees of freedom of the valence electron system are suppressed. Physically this
situation can be realized in a fully saturated ferromagnet.

The final and initial states of the system are determined by a Hamiltonian $H_0$ that
describes the electrons around the Fermi-level. The generic form of the full
many-body Hamiltonian is
\begin{eqnarray}
H_0 =   \sum_{i,j} t_{ij}  (c^{\tiny \dagger}_i c^{\phantom{\tiny
\dagger}}_j + c^{\tiny \dagger}_j c^{\phantom{\tiny \dagger}}_i) +
c^{\tiny \dagger}_i c^{\phantom{\tiny \dagger}}_i V_{ij} c^{\tiny
\dagger}_j c^{\phantom{\tiny \dagger}}_j, \label{eq:H0}
\end{eqnarray}
where $i$ and $j$ denote lattice sites with lattice vectors ${\bf
R}_i$ and ${\bf R}_j$. Note that the sum is over each pair $i,j$
once, with $i,j$ ranging from 1 to $N$, where $N$ is the number of
sites in the system. The hopping amplitudes of the valence
electrons are denoted by $t_{ij}$ and the $c/c^{\tiny
\dagger}$-operators annihilate/create such electrons. The Coulomb
interaction between valence electrons is  $V_{ij}= V_{|{\bf
R}_i-{\bf R}_j|}$, as the Coulomb interaction only depends on the
relative distance between two particles.

The intermediate states are eigenstates of the Hamiltonian $H_{\rm int}= H_0 + H_c$, where $H_c$
accounts for the Coulomb coupling between the intermediate state core-hole and the valence electrons:
\begin{eqnarray}
H_c = \sum_{i,j} s^{\phantom{\tiny \dagger}}_i s^{\tiny \dagger}_i
V^{c}_{ij} c^{\tiny \dagger}_j c^{\phantom{\tiny \dagger}}_j,
\label{eq:Hcore}
\end{eqnarray}
where $s_i$ creates a core-hole on site $i$.
We assume that the core-hole is fully localized
and has no dispersion.
We will see shortly that this leads to major simplifications in the
theoretical treatment of indirect RIXS. The core-hole -- valence electron interaction is attractive:  $V^{c} < 0$.
The dipole operators are given by
\begin{eqnarray}
\hat{O} = \sum_{i}  e^{-i {\bf q}_{\rm in}\cdot{\bf R}_i }
s^{\phantom{\tiny \dagger}}_i  p^{\tiny \dagger}_i + e^{i {\bf
q}_{\rm out}\cdot{\bf R}_i } s^{\tiny \dagger}_i
p^{\phantom{\tiny \dagger}}_i +h.c.,
\end{eqnarray}
where $p^{\tiny \dagger}$ creates a photo-excited electron in a $4p$ state and $h.c.$ denotes the Hermitian conjugate of both terms.

\subsection{Short Lifetime Approximation: Algebraic Form}
In order to calculate the cross section, we need to evaluate the operator $(H_{\rm int})^l = (H_0 + H_c)^l$ in equation (\ref{eq:Mterms}). A direct evaluation of this operator is complicated by the fact that $[H_0,H_c] \neq 0$. We therefore proceed by approximating $H_{\rm int}^l$ with a series that contains the leading terms to the scattering cross section for both strong and weak core-hole potentials, if the lifetime is short. After that we will do a full re-summation of that series. This approximation is central to the results in this paper.

Expanding $(H_0 + H_c)^l$ gives a series with $2^l$ terms:
\begin{align}
	H_{\text{int}}^l = &H_c^l + \sum_{n=0}^{l-1} H_c^n H_0^{\phantom{n}} H_c^{l-n-1} + \dots + \nonumber \\
	&\sum_{n=0}^{l-1} H_0^n H_c^{\phantom{n}} H_0^{l-n-1} + H_0^l.
\end{align}
Using $H_0 \hat{O} \ket{\textsc{i}} = \hat{O} H_0 \ket{\textsc{i}} \equiv 0$, this series reduces to
\begin{align}
	H_{\text{int}}^l \hat{O} \ket{\textsc{i}} = [ H_c^l + &\sum_{n=0}^{l-2} H_c^n H_0^{\phantom{n}} H_c^{l-n-1} + \dots + \nonumber \\
	& H_0^{l-1} H_c^{\phantom{n}} ] \hat{O}\ket{\textsc{i}}.
\end{align}
Using in addition that $\bra{\textsc{f}} \hat{O} H_0 = \bra{\textsc{f}} H_0 \hat{O} = E_{\textsc{f}} \bra{\textsc{f}} \hat{O}$, we find
\begin{align}
	\bra{\textsc{f}} \hat{O} &H_{\text{int}}^l \hat{O} \ket{\textsc{i}} = \bra{\textsc{f}} \hat{O} [ H_c^l + E_{\textsc{f}}^{\phantom{n}} H_c^{l-1} + \nonumber \\
	&\sum_{n=1}^{l-2} H_c^n H_0^{\phantom{n}} H_c^{l-n-1} + \dots + E_{\textsc{f}}^{l-1} H_c ] \hat{O}\ket{\textsc{i}}.
	\label{eq:exp_lhs}
\end{align}

For strong core-hole potentials, the leading term of $H_{\rm int}^l$ is $H_c^l$. Corrections to this term contain at least one factor of $H_0$ and are therefore smaller by a factor of at least $t/V^c$. For weak core-hole potentials, the term $H_0^l$ vanishes because $[H_0,\hat{O}]=0$. The leading term for this limit therefore is $E_{\textsc{f}}^{l-1} H_c$. Correction terms contain at least two factors of $H_c$, which make them at least a factor of $V^c/t$ smaller.

Now we consider the approximation
\begin{equation}
	H_{\rm int}^l \hat{O} \ket{\textsc{i}} \simeq \sum_{m=0}^{l} H_{0}^m H_c^{l-m} \hat{O} \ket{\textsc{i}}.
	\label{eq:H_expand}
\end{equation}
It can be seen that the leading order terms for both strong $(m=0)$ and weak $(m=l-1)$ core-hole potentials are included in the sum. The other terms are included only for mathematical convenience lateron; they can be neglected if we consider either limit. Note that the $m=l$ term in eq.~(\ref{eq:H_expand}) is $0$, so that it can be removed from the sum. After performing the same manipulations as above, we obtain
\begin{align}
	\bra{\textsc{f}} \hat{O} &\sum_{m=0}^{l-1} H_{0}^m H_c^{l-m} \hat{O} \ket{\textsc{i}} = \sum_{m=0}^{l-1} E_{\textsc{f}}^m \bra{\textsc{f}} \hat{O} H_c^{l-m} \hat{O} \ket{\textsc{i}} \nonumber \\
	&=\bra{\textsc{f}} \hat{O} [ H_c^l + E_{\textsc{f}} H_c^{l-1} + \dots + E_{\textsc{f}}^{l-1} H_c ] \hat{O} \ket{\textsc{i}}.
	\label{eq:exp_rhs}
\end{align}
Comparing eqs.~(\ref{eq:exp_lhs}) and (\ref{eq:exp_rhs}), it can be seen that the approximation (\ref{eq:H_expand}) is exact in the limit of both strong and weak core-hole potentials.

\subsection{Short Life-time Approximation: Graphical Representation.}
\begin{figure}
\includegraphics[width=0.9\columnwidth]{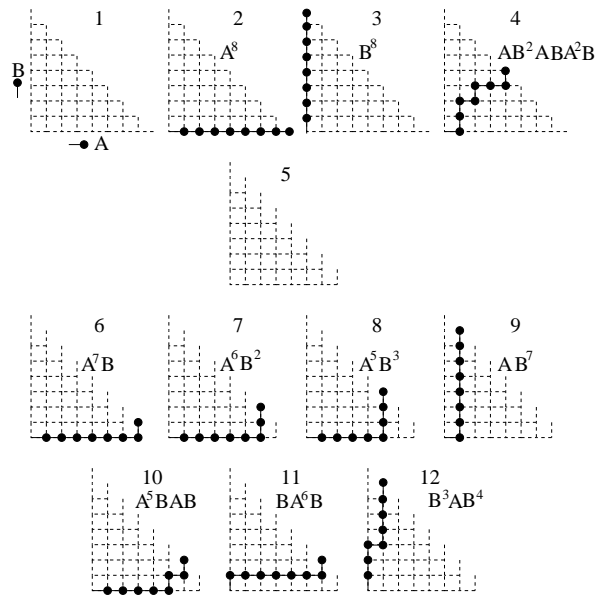} 
\caption{Graphical representation of the expansion of $(A+B)^l$, where $A=H_0$ and $B=H_c$ are two non-commuting operators. In this example $l=8$.}
\label{fig:expand}
\end{figure}

We can also represent the series expansion and its approximation graphically (Fig.~\ref{fig:expand}). When we expand $(A+B)^l$, where $A$ and $B$ are non-commuting operators, each term in the series corresponds to a graph on the grid of Fig.~\ref{fig:expand}.1. Each graph occurs only once and can be constructed by starting at the lower left corner of the grid and moving either to the right, representing an $A$, or up, representing a $B$. At the next vertex a new move (right or up) is made. We perform this procedure $l$ times and in this way we can obtain $2^l$ distinct graphs, each corresponding to a term in the expansion of $(A+B)^l$. For example moving $l$ times to the right represents the term $A^l$ and moving $l$ times up corresponds to $B^l$, see Fig.~\ref{fig:expand}.2 and ~\ref{fig:expand}.3. All other terms in the series can be constructed by moving up and right a different number of times and in different order. As we consider a fixed value of $l$ ($l=8$ in Fig.~\ref{fig:expand}), all graphs must end on the diagonal of the triangle that forms the grid.
In the series for $(H_0+H_c)^l \hat{O} \ket{\textsc{i}}$ ($H_0=A$ and $H_c = B$) we have the simplification that terms ending with $H_0$ acting on the groundstate give zero. These terms can thus be removed from the expansion. The graphs for this expansion now live on a reduced grid where the horizontal grid-lines at the diagonal of the triangle are absent, see Fig.~\ref{fig:expand}.5: these represent all terms ending on $A$.

In Fig.~\ref{fig:expand} we also represent the approximate series of the r.h.s. of Eq. ~(\ref{eq:H_expand}). Graphically this sum corresponds to the set of graphs on the reduced grid of Fig.~\ref{fig:expand}.5, with either one kink (Fig.~\ref{fig:expand}.6, \ref{fig:expand}.7, \ref{fig:expand}.8, \ref{fig:expand}.9) or without kinks (Fig.~\ref{fig:expand}.2, \ref{fig:expand}.3). 
Thus, in our approximation in Eq. ~(\ref{eq:H_expand}) of the exact series for  $(H_0+H_c)^l $ we neglect all graphs with two or more kinks (Fig.~\ref{fig:expand}.4, \ref{fig:expand}.10, \ref{fig:expand}.11, \ref{fig:expand}.12). In the limit of either very $A$ or very large $B$, the graphs that we neglect correspond to sub-leading order corrections. When $A$ is largest then the leading terms are, first, graph \ref{fig:expand}.2, which is however zero because it ends on $A$. The leading term is therefore of the order $A^7$ and shown in graph \ref{fig:expand}.6. Other higher order terms are shown in the graphs \ref{fig:expand}.7, \ref{fig:expand}.8, \ref{fig:expand}.10 and \ref{fig:expand}.11. The last two graphs are neglected in our approximate expansion. In case $B$ is dominating, the leading term is $B^8$, graph \ref{fig:expand}.3, and next to leading is graph \ref{fig:expand}.9, with $B^7$. The highest order terms that are neglected in our approximate series are of the type shown in graph \ref{fig:expand}.12.

\subsection{Re-summation of Series for Scattering Cross Section}
In order to obtain $M_l$ and from there the scattering amplitude
$A_{\textsc{fi}}$ and finally the scattering cross-section, in
Eq.(\ref{eq:exp_rhs}) we need to evaluate expressions of the kind
\begin{eqnarray}
H_c^n \hat{O}| \textsc{i} \rangle = H_c^{n-1}  \sum_{i,l,j}
s^{\phantom{\tiny \dagger}}_l s^{\tiny \dagger}_l V^{c}_{lj}
c^{\tiny \dagger}_j c^{\phantom{\tiny \dagger}}_j e^{-i {\bf
q}_{\rm in}\cdot{\bf R}_i } s^{\phantom{\tiny \dagger}}_i p^{\tiny
\dagger}_i | \textsc{i} \rangle. \label{eq:H_c^n,n-1}
\end{eqnarray}
In the initial state no core-hole is present: just one core-hole
is created by the dipole operator. We therefore have that $
s^{\phantom{\tiny \dagger}}_l s^{\tiny \dagger}_l
s^{\phantom{\tiny \dagger}}_i | \textsc{i} \rangle = \delta_{l,i}
s^{\phantom{\tiny \dagger}}_i s^{\tiny \dagger}_i
s^{\phantom{\tiny \dagger}}_i | \textsc{i} \rangle= \delta_{l,i}
s^{\phantom{\tiny \dagger}}_i | \textsc{i} \rangle$. Inserting
this in equation~(\ref{eq:H_c^n,n-1}), we obtain
\begin{eqnarray}
H_c^n \hat{O}| \textsc{i} \rangle = H_c^{n-1}  \sum_i  e^{-i {\bf
q}_{\rm in}\cdot{\bf R}_i } s^{\phantom{\tiny \dagger}}_i p^{\tiny
\dagger}_i  \sum_j  V^{c}_{ij} c^{\tiny \dagger}_j
c^{\phantom{\tiny \dagger}}_j  | \textsc{i} \rangle
\end{eqnarray}
and by recurrence
\begin{eqnarray}
H_c^n \hat{O}| \textsc{i} \rangle =
 \sum_i  e^{-i {\bf q}_{\rm in}\cdot{\bf R}_i } s^{\phantom{\tiny \dagger}}_i  p^{\tiny \dagger}_i  \left[\sum_j  V^{c}_{ij} c^{\tiny \dagger}_j c^{\phantom{\tiny \dagger}}_j \right] ^n | \textsc{i} \rangle
\label{eq:H_c^n}
\end{eqnarray}

Let us for the moment consider the strong core-hole potential limit and keep in the expansion Eq.(\ref{eq:H_expand}) only the term $m=0$.
Inserting the results above in Eq.(\ref{eq:Mterms}), we find that
\begin{eqnarray}
M_l (V^{c}  \gg t)= \frac{1}{\Delta^l} \langle \textsc{f} |
\sum_{i} e^{i {\bf q} \cdot{\bf R}_i } \left[ \sum_{j} V^{c}_{ij}
c^{\tiny \dagger}_j c^{\phantom{\tiny \dagger}}_j \right]^l  |
\textsc{i} \rangle, \label{eq:M_strong}
\end{eqnarray} where the transfered momentum
${\bf q} \equiv {\bf q}_{\rm out} - {\bf q}_{\rm in}$.

The first important observation is that the term $l=0$ does not
contribute to the inelastic X-ray scattering intensity because
$M_0=  \langle \textsc{f} | \sum_{i} e^{i {\bf q} \cdot{\bf R}_i }
| \textsc{i} \rangle = N \delta_{\bf q,0}
\delta_{\textsc{f},\textsc{i}}$, which only contributes to the
elastic scattering intensity at ${\bf q}={\bf 0}$ and other
multiples of the reciprocal lattice vectors. From inspection of
equation (\ref{eq:M_l}) we see immediately that the $l=0$ term
actually vanishes irrespective of the strength of the core-hole
potential. This is of relevance when we consider the scattering
cross section in the so-called "fast-collision
approximation"~\cite{Veenendaal96}.  This approximation
corresponds to the limit where the core-hole life time broadening
is the largest energy scale in system ($\Gamma \rightarrow \infty$
or, equivalently, $Im [ \Delta ] \rightarrow -\infty$). In this
limit only the $l=0$ term contributes to the indirect RIXS
amplitude and
 the resonant inelastic signal vanishes. In any theoretical treatment of indirect resonant scattering one therefore
needs to go beyond the fast-collision approximation.

Physically this vanishing of spectral weight  is ultimately due to
an interference effect. If we study a process in which we start
from the initial state and reach a certain final state, we need to
consider all different possible paths for this excitation --
de-excitation process. When the core-hole broadening is very large
we can reach the final state via any intermediate state and in
order to obtain the scattering amplitude we thus add up coherently
the contributions of all intermediate states. We then obtain $A =
\sum_n \langle \textsc{f} | n \rangle \langle n|
\textsc{i}\rangle$. When the set of intermediate states that we
sum over is complete (which by definition is the case when $\Gamma
\rightarrow \infty$ ), this leaves us with $A=\langle \textsc{f} |
\textsc{i}\rangle$ which is, because of the orthogonality of
eigenstates, only non-zero when the initial and final state are
equal --hence only when the scattering is elastic.

The second observation is that $M_l$ is a $2^l$-particle correlation function. If we measure far away from resonance,
where $|Re[\Delta]| \gg 0$, the scattering cross section is dominated by the $l=1$, two-particle, response function. When
the incoming photon energy approaches the resonance, gradually the four, six, eight etc. particle
response functions add more and more spectral weight to the inelastic scattering amplitude. Generally these
multi-particle response functions interfere. We will show, however, that in the local core-hole approximation the multi-particle
correlation functions in expansion (\ref{eq:H_expand}) collapse onto the dynamic two-particle (charge-charge) and
four-particle (spin-spin) correlation function.

\subsection{Local core-hole potentials}
In hard X-ray electron spectroscopies one often makes the approximation that the core-hole potential is local. This corresponds
to the widely used Anderson impurity approximation in the theoretical analysis of e.g. X-ray absorption and photo-emission,
introduced in Refs.~\cite{Gunnarsson83,Kotani85,Zaanen86}.
This approximation is reasonable as the Coulomb potential is certainly largest on the atom where the core-hole is
located.

In the present case, moreover, we can consider the potential generated by both the localized core-hole and photo-excited electron at the same time. As this exciton is a neutral object, its monopole contribution to the potential vanishes for distances larger than the exciton radius. The multi-polar contributions that we are left with in this case are generally small and drop off quickly with distance.

We insert a local core-hole potential $V^{c}_{ij}=U\delta_{ij}$  in our equations and aim to re-sum the approximate series expansion in Eq.(\ref{eq:H_expand}) for arbitrary values of the local core-hole potential. We find from  Eq.(\ref{eq:H_c^n}) that
\begin{eqnarray}
H_c^n \hat{O}| \textsc{i} \rangle =
 \sum_i  e^{-i {\bf q}_{\rm in}\cdot{\bf R}_i } s^{\phantom{\tiny \dagger}}_i  p^{\tiny \dagger}_i   U^n  [c^{\tiny \dagger}_i c^{\phantom{\tiny \dagger}}_i]^n | \textsc{i}
 \rangle \label{eq:H_c^nloc}
\end{eqnarray}
Using that for fermions $[c^{\tiny \dagger}_i c^{\phantom{\tiny
\dagger}}_i]^n = c^{\tiny \dagger}_i c^{\phantom{\tiny
\dagger}}_i$, we obtain for our spinless fermions
\begin{eqnarray}
M_l^{sf}= \frac{1}{\Delta^l}  \langle \textsc{f} | \sum_{i} e^{i
{\bf q} \cdot{\bf R}_i }  c^{\tiny \dagger}_i c^{\phantom{\tiny
\dagger}}_i  | \textsc{i} \rangle \sum_{m=0}^{l-1} E_\textsc{f}^m
U^{l-m}.
\end{eqnarray}
The sum over $m$ can easily be performed:
\begin{eqnarray}
\sum_{m=0}^{l-1} E_\textsc{f}^m U^{l-m}= U^l \sum_{m=0}^{l-1}
(E_\textsc{f}/U)^m =
 \frac{U^l- E_\textsc{f}^l}{1- E_\textsc{f}/U}
\end{eqnarray}
and we obtain
\begin{eqnarray}
M^{sf}_l =  \frac{1}{\Delta^l}
\frac{U^l-{E}_\textsc{f}^l}{1-E_\textsc{f}/U} \langle \textsc{f} |
\sum_{i} e^{i {\bf q} \cdot{\bf R}_i }  c^{\tiny \dagger}_i
c^{\phantom{\tiny \dagger}}_i  | \textsc{i} \rangle.
\label{eq:M_local}
\end{eqnarray}
Using that $\sum_{i} e^{i {\bf q} \cdot{\bf R}_i }  c^{\tiny
\dagger}_i c^{\phantom{\tiny \dagger}}_i  = \sum_{\bf k} c^{\tiny                                     
\dagger}_{{\bf k}-{\bf q}} c^{\phantom{\tiny \dagger}}_{\bf k}
\equiv \rho_{\bf q}$  is the density operator, we have to perform
the sum over $l$ in equation~(\ref{eq:A_fi}). The $l=0$ term is
zero, as we discussed above, so that the scattering amplitude is
\begin{eqnarray}
A_{\textsc{fi}} = \frac{ \omega_{\rm res} } {\Delta}
\sum_{l=1}^{\infty} M_l. \label{eq:A_fi}
\end{eqnarray}
Using
\begin{eqnarray}
\sum_{l=1}^{\infty} (U/\Delta)^l-(E_\textsc{f}/\Delta)^l = \Delta
\frac{U-E_\textsc{f}}{(\Delta-U)(\Delta-E_\textsc{f})}
\end{eqnarray}
we finally find that the indirect resonant inelastic scattering amplitude for spinless fermions is
\begin{eqnarray}
A_{\textsc{fi}}^{sf}  =    P_1(\omega,U) \langle \textsc{f}|
\rho_{\bf q} | \textsc{i} \rangle, \label{eq:A_spinless}
\end{eqnarray}
where  the resonant enhancement factor is $P_1(\omega,U)  \equiv U
\omega_{\rm res}  [(\Delta-U)(\Delta-\omega)] ^{-1}$ and
$\omega=E_\textsc{f}$. For spinless fermions with a local core-hole
potential the scattering cross section thus turns out to be the
density response function --a two-particle correlation function--
with a resonant prefactor $P_1(\omega)$ that depends on the loss
energy $\omega$, the resonant energy ${\omega}_{\rm res}$,  on the
distance from resonance $\omega_{\rm in}(=Re[ \Delta ]$), on the
core-hole potential $U$ and on the core-hole life time broadening
$\Gamma(=-Im [ \Delta ])$.  We see that the resonant enhancement
is largest when the energy of the incoming photons is either equal
to the core-hole potential ($\omega_{\rm in}=U$) or to the loss
energy ($\omega_{\rm in}=\omega$), which one could refer to as a
`final-state resonance'.

The density response function is related to the dielectric function
$\epsilon({\bf q},\omega)$ and the dynamic structure factor $S({\bf q},\omega)$~\cite{Mahan90}, so that we obtain for the resonant scattering cross section
\begin{eqnarray}
\left. \frac{ {\rm d}^2 \sigma} { {\rm d} \Omega {\rm d} \omega}  \right|_{res}^{sf}
&\propto& -|P_1(\omega)|^2  \ {\rm Im} \ \left[ \frac{1}{V_{\bf q} \epsilon({\bf q},\omega)} \right] \nonumber \\
&\propto& |P_1(\omega)|^2  \ S({\bf q},\omega),
\end{eqnarray}
for a fixed  value of the core-hole potential $U$. $V_{\bf q}$ is the Fourier transform of the Coulomb potential. For weak core-hole potentials the total scattering intensity is proportional to $U^2$ and for strong core-hole potentials, where $|U| \gg \Gamma$, the scattering intensity at resonance ($\omega_{\rm in}=0$) is to first order independent of the strength of the core-hole potential. Far away from the edge, however, where $|\omega_{\rm in}| \gg |U|$, the scattering intensity is again proportional to $U^2$, just as for weak core-hole potentials.
Integrating  $|P_{1}(\omega)|^2$ over all incoming photon energies, we obtain the integrated inelastic intensity at fixed loss energy $\omega$ and momentum $\bf q$
\begin{eqnarray}
\int_{-\infty}^{\infty} \left. d \omega_{\rm in}  \frac{ {\rm d}^2 \sigma} { {\rm d} \Omega {\rm d} \omega}  \right|_{res}^{sf}   \propto
\frac{2 \pi U^2 \omega_{\rm res}^2 }{\Gamma (4 \Gamma^2 +(U-\omega)^2)}  \ S({\bf q},\omega).
\end{eqnarray}
It seems that the resonant enhancement factor of the integrated
intensity has a maximum when the loss energy is equal to the
core-hole potential. However, the core-hole potential is
attractive and therefore lower than zero, and the loss energy
$\omega$ is by definition greater than zero.  So the integrated
intensity is maximal at energy loss $\omega=0$.

\section{Indirect RIXS for spinless fermions: finite T}
In this section, we generalize the previous calculation
to the case of finite temperature. The starting point is as before
\begin{equation}
\left. \frac{ {\rm d}^2 \sigma} { {\rm d} \Omega {\rm d} \omega}
\right|_{res} \propto \frac{1}{Z}\sum_{\textsc{i}}
\sum_{\textsc{f}}  | A_{\textsc{fi}} |^2 \ \delta
(\omega-\omega_{\textsc{fi}}) e^{-\beta E_\textsc{i}},
\label{eq:KramersT}
\end{equation}
where $Z= \sum_{\textsc{i}} e^{-\beta E_\textsc{i}}$ is the
partition function and $\beta=1/k_{\rm B} T$. Equation
(\ref{eq:KramersT}) represents the statistical average over all
the initial states $| \textsc{i} \rangle$, where now the more
general relation $H_0 | \textsc{i} \rangle
=E_{\textsc{i}} | \textsc{i} \rangle $ holds. 

We expand the scattering amplitude $A_{\textsc{f} \textsc{i}}$, using again the ultrashort life time of  the core-hole as in 
Eq. (\ref{eq:A_fi}). We are left with
the evaluation of the operator $(H_{\rm int})^l$. We proceed by
expanding it in the following way:
\begin{eqnarray}
(H_{\rm int})^l \hat{O} | \textsc{i} \rangle
&=& (H_0+H_c)^l   \hat{O} | \textsc{i} \rangle     \nonumber \\
&\simeq& \sum_{n=0}^{l-1} \sum_{m=0}^{l-n-1} (H_{0})^{m} (H_c)^{l-m-n} (H_{0})^{n} \hat{O}|
\textsc{i} \rangle, \label{eq:H_expandT}
\end{eqnarray}
where we neglected the term $H_0^l$, as it will not contribute to the inelastic scattering cross section. 
This approximation reproduces the correct leading order terms, which represent the strong and weak
coupling case, respectively. Moreover, it is a generalization of
(\ref{eq:H_expand}), that takes into account that the initial
state is no longer the ground state so that $H_0 | \textsc{i}
\rangle =E_{\textsc{i}} | \textsc{i} \rangle $. In our graphical
representation, with respect to the $T=0$ case, it corresponds to
retain all the additional terms, having more than one kink, that
start and finish with a horizontal step. In doing this, we are
neglecting again
the sub-leading order terms $H_c^{l-1-n} H_0 H_c^n$.

After inserting expansion (\ref{eq:H_expandT}) in the expression
(\ref{eq:Mterms}) for $M_{l}$, we finally have to evaluate
\begin{eqnarray}
\langle \textsc{f} | \hat{O} \sum_{n,m} (H_{0})^{m} (H_c)^{l-m-n}
(H_{0})^{n} \hat{O}| \textsc{i} \rangle =     \nonumber \\
\sum_{n,m} E_\textsc{f}^m E_\textsc{i}^n \langle \textsc{f} |
\hat{O} H_c^{l-m-n} \hat{O}| \textsc{i} \rangle.
\end{eqnarray}
In the local core-hole approximation, we can resum this
approximate series expansion. By using the results of
Eqs.~(\ref{eq:H_c^nloc}), we obtain for spinless
fermions
\begin{eqnarray}
M_l^{sf}= \frac{1}{\Delta^l}  \langle \textsc{f} | \rho_{\bf q} |
\textsc{i} \rangle U^{l}\sum_{n,m} E_\textsc{f}^m E_\textsc{i}^n
U^{l-m-n}.
\end{eqnarray}
By performing the sums over $n$ and $m$
\begin{eqnarray}
U^{l}\sum_{n,m} E_\textsc{f}^m E_\textsc{i}^n U^{l-m-n}= \nonumber \\
U^{l}\sum_{n=0}^{l-1} (E_\textsc{i}/U)^n \sum_{m=0}^{l-n-1} (E_\textsc{f}/U)^n,
\end{eqnarray}
and after summing over $l$, we finally obtain
\begin{eqnarray}
A_{\textsc{fi}}^{sf}  =    P_1(E_\textsc{f},U)
\frac{\Delta}{\Delta-E_{\textsc{i}}} \langle \textsc{f}| \rho_{\bf
q} | \textsc{i} \rangle . \label{eq:A_spinlessT}
\end{eqnarray}
This equation clearly shows that one of the main effects of finite
temperature is to modify the resonant enhancement factor,
nevertheless preserving the same structure for the scattering
amplitude.

At this point we observe that at resonance $|\Delta|=\Gamma$, which is of the order of electron volts and thus several orders of
magnitude larger than $E_{\textsc{i}}$, even at high temperature. 
This allows us to approximate the prefactor in Eq.~(\ref{eq:A_spinlessT}) as
\begin{eqnarray}
&&P_1(E_\textsc{f},U) \frac{\Delta}{\Delta-E_{\textsc{i}}} \simeq     \nonumber \\
&&P_1(\omega,U)(1+\frac{E_\textsc{i}}{\Delta-\omega}+....)(1+\frac{E_\textsc{i}}{\Delta}+....).
\label{eq:noname}
\end{eqnarray}
At the lowest order in $E_\textsc{i}/\Gamma$, the prefactor is not
modified by $T$ at all, hence we conclude that  the major
modifications to the cross section are induced by thermal
averaging of the correlation function. After integrating over all
the incoming photon energies, we get the following approximate
expression for the thermal average of the inelastic
intensity at  loss energy $\omega$ and momentum $\bf q$ 

\begin{eqnarray}
\left. \frac{ {\rm d}^2 \sigma} { {\rm d} \Omega {\rm d} \omega}  \right|_{res, {\rm T}}
&\propto& |P_1(\omega)|^2 
\left\langle S({\bf q},\omega) \right\rangle_{\rm T}.
\end{eqnarray}

In this expression the temperature dependence is entirely due to the temperature dependence 
of $S({\bf q},\omega)$. The pre-factor is in leading order temperature independent.  At finite
temperatures energy gain scattering can occur: the photon
can gain an energy of the order of $k_BT$ from the system, which corresponds to negative energy loss.

\section{Fermions with spin}
\begin{figure}
\includegraphics[width=0.6\columnwidth]{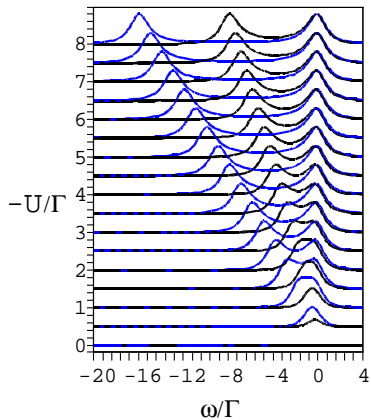}
\caption{ Prefactors  to the scattering intensity $|P_1(U)|^2$ (black) and $|P_2(U)|^2$ (blue)  at fixed loss
energy $\omega$ as a function of incoming photon energy $\omega_{\rm in}/\Gamma$ for different values
of the local core-hole potential $U/\Gamma$}
\label{fig:resonance}
\end{figure}

We generalize the calculation above to the situation where the electrons have an additional spin degree of freedom. In the Hamiltonians ~(\ref{eq:H0},\ref{eq:Hcore}) we now include a spin index $\sigma$ (with  $\sigma= \uparrow$ or $\downarrow$)
to the annihilation and creation operators: $c_i \rightarrow c_{i \sigma}$ and $c_j \rightarrow c_{j \sigma^{\prime}}$
and sum over these indices, taking into account that the hopping part of the Hamiltonian is diagonal in the spin variables.
In order to re-sum the series in equation~ (\ref{eq:H_expand}) we now need to evaluate expansions of the number
operators of the kind $(n_{\uparrow}+n_{\downarrow})^l$.
Using
\begin{eqnarray}
(n_{\uparrow}+n_{\downarrow})^l &=& n_{\uparrow} + n_{\downarrow}  +
n_{\uparrow}n_{\downarrow} \sum_{p=1}^{l-1} \left( \begin{array}{c} l \\ p \end{array} \right)  \nonumber \\
&=& n_{\uparrow} + n_{\downarrow}   + (2^l-2) n_{\uparrow}n_{\downarrow} ,
\end{eqnarray}
for $l >0$, we obtain
\begin{equation}
A_{\textsc{fi}}= \langle \textsc{f}| P_1(\omega)
[\rho^{\phantom{\uparrow}}_{\bf q} - 2\rho^{\uparrow \downarrow}_
{\bf q} ]+ 2 P_2 (\omega) \rho^{\uparrow \downarrow}_{\bf q}|
\textsc{i} \rangle,
\end{equation}
with $P_2(\omega,U) = P_1(\omega,2U)/2 $ and $\rho^{\uparrow
\downarrow}_{\bf q}  \equiv \sum_{i}  e^{i {\bf q} \cdot {\bf
R}_i}  n_{i \uparrow} n_{i \downarrow} $. We see that in the case
that each site can only be occupied by at most one valence
electron, this equation immediately reduces to
Eq.~(\ref{eq:A_spinless}) with $\rho^{\phantom{\uparrow}}_{\bf q}
= \rho^{\uparrow} _{\bf q} + \rho^{\downarrow}_ {\bf q} $. The two
terms in the scattering amplitude  can also be written in terms of
density and spin operators. Using $  (n_{i \uparrow} -n_{i
\downarrow})^2 =(2S^z_i)^2 =  \frac{4}{3}  {\bf S}_i^2$, we obtain
$\rho^{\phantom{\uparrow}}_{\bf q} - 2\rho^{\uparrow \downarrow}_
{\bf q} =   {\bf S} _{\bf q}^2$, where we  introduce the
longitudinal spin density correlation function  ${\bf S}_{\bf q}^2
\equiv \frac{1}{S(S+1)}\sum_{\bf k} {\bf S_{k+q}} \cdot {\bf
S_{-k} } $. In terms of these correlation functions the scattering
amplitude for spinfull fermions is
\begin{equation}
A_{\textsc{fi}}= [P_1(\omega)- P_2(\omega)] \langle \textsc{f}|
{\bf S}_{\bf q}^2 | \textsc{i} \rangle   + P_2(\omega)  \langle
\textsc{f}|\rho_{\bf q}  | \textsc{i} \rangle,
\end{equation}
Clearly the contributions to the scattering rate from the dynamic longitudinal spin correlation function and the density correlation function need to be treated on equal footing as they interfere~\cite{Devereaux03-Kondo01,Note2}.
Moreover, the spin and charge correlation functions have different resonant enhancements, see Fig.~\ref{fig:resonance}. For instance when $Re[\Delta]=U$, the scattering amplitude is dominated by $P_1(\omega)$ and hence by the longitudinal spin response function. At incident energies where $Re[\Delta]=2U$, on the other hand, $P_2(\omega)$ is resonating so that the contributions to the inelastic scattering amplitude of charge and spin are approximately equal.

\section{Multi-band systems}
Let us consider systems with more than one band and take as an explicit example a transition metal with a $3d$ and a $4s$ band. The Coulomb attraction between the $1s$ core-hole and an electron in the $3d$ state ($U_d$) is much larger than the interaction with a $4s$ electron ($U_s$). Neglecting spin degrees of freedom we would naively expect that the indirect RIXS response in the two-band system is simply the sum of the responses of the two individual electronic systems, with possible interference between the two scattering channels: we expect the scattering amplitude to be equal to
\begin{equation}
 A_{\textsc{fi}}^{s+d} =P_1(\omega,U_d) \langle \textsc{f}| \rho^d_{\bf q}    | \textsc{i} \rangle   + P_1(\omega,U_s)  \langle \textsc{f}|\rho^s_{\bf q}  | \textsc{i} \rangle.
\end{equation}
However, already from the calculation for the spinfull fermions we know that the situation should be more complicated, as in that case the full response function is not just the sum of the two response functions for spinless fermions. The point is that when both a $3d$ and $4s$ electron screen the core-hole, the intermediate state is at a lower energy (at $\omega_{\rm in} = U_d+U_s$) compared to the situation where only a single $d$/$s$ electron screens the core-hole (with a resonance at $\omega_{\rm in} =U_d$/$U_s$, respectively.) In the situation that both electrons screen the core-hole, the resonance therefore appears at a different incoming photon energy.

According to Eq.(\ref{eq:H_c^n}), we now need to evaluate expressions of the sort $(U_d n^d+U_s n^s)^l$ for $l>0$. After using the binomial theorem and summing the resulting series, we obtain
 \begin{eqnarray}
(U_d n^d+U_s n^s)^l &=&  U_d^l n^d+U_s^l n^s \nonumber \\
&+& n^d n^s  [ (U_d+U_s)^l-U_d^l-U_s^l],
\end{eqnarray}
which leads to a scattering amplitude
\begin{eqnarray}
 A_{\textsc{fi}}^{sd} &=&A_{\textsc{fi}}^{s+d} + [P_1(\omega,U_d+U_s) \nonumber \\
&-& P_1(\omega,U_d) - P_1(\omega,U_s) ] \langle \textsc{f}|
\rho^{ds}_{\bf q}    | \textsc{i} \rangle,
\end{eqnarray}
where $\rho^{ds}_{\bf q}  \equiv \sum_{i}  e^{i {\bf q} \cdot {\bf R}_i}  n_i^d n_i^s $.

\section{ Conclusions}
On the basis of the ultrashort life-time of the core-hole in the intermediate state 
we presented a series expansion of the indirect resonant inelastic
X-ray scattering amplitude, which is asymptotically exact for both
small and large local core-hole potentials. This algebraic series is also given in a graphical representation.
By re-summing the terms in the series, we find the dynamical charge and spin
correlation functions that are measured in RIXS. The resonant pre-factor is only weakly temperature dependent.
We have also derived a sum-rule for the total scattering intensity and considered RIXS in both single and multi-band systems.
On the basis of our results, the charge and spin structure factor of e.g. Hubbard-like model Hamiltonians can be used to be directly compared to the experimental RIXS spectra.

\section{Acknowledgments}
We thank Michel van Veenendaal and John P. Hill for stimulating our
interest in the theory of indirect resonant inelastic X-ray
scattering, for intense discussions, for critical reading of the
manuscript and for their generous hospitality. We thank  Stephane
Grenier, Young-June Kim,  Philip Platzman and George Sawatzky for
fruitful discussions. We gratefully acknowledge support from the
Argonne National Laboratory Theory Institute, Brookhaven National
Laboratory (DE-AC02-98CH10996), the Dutch Science Foundation FOM.

\end{document}